# Unfolding the fast neutron spectra of a BC501A liquid scintillation detector using GRAVEL method

CHEN YongHao[1,2*], CHEN XiMeng[1], LEI JiaRong[2], AN Li[2], ZHANG XiaoDong[1] SHAO JianXiong[1], ZHENG Pu[2] & WANG XinHua[2]

[1]*School of Nuclear Science and Technology, Lanzhou University, Lanzhou 730000, China*
[2]*Institute of Nuclear Physics and Chemistry, China Academy of Engineering Physics, Mianyang 621900, China*

Accurate knowledge of the neutron energy spectra is useful in basic research and applications. The overall procedure of measuring and unfolding the fast neutron energy spectra with BC501A liquid scintillation detector is described. The recoil proton spectrum of [241]Am-Be neutrons was obtained experimentally. With the NRESP7 code, the response matrix of detector was simulated. Combining the recoil proton spectrum and response matrix, the unfolding of neutron spectra was performed by GRAVEL iterative algorithm. A MatLab program based on the GRAVEL method was developed. The continuous neutron spectrum of [241]Am-Be source and monoenergetic neutron spectrum of D-T source have been unfolded successfully and are in good agreement with their standard reference spectra. The unfolded [241]Am-Be spectrum are more accurate than the spectra unfolded by artificial neural networks in recent years.

**neutron spectroscopy, GRAVEL algorithm, unfolding of neutron spectra, BC501A liquid scintillator, NRESP7**

## 1. Introduction

Accurate knowledge of the neutron energy spectra is of great research importance in many basic research and applications, such as in nuclear nonproliferation, international safeguard, nuclear material control, nation security and counterterrorism [1-2]. In addition, the accurate unfolding of neutron spectra increases the sensitivity of assays performed on various nuclear materials [3]. BC501A liquid scintillation detectors are widely used for the fast neutron spectroscopy because of their good linearity, excellent n-γ discrimination property and high light output [4-6]. However, the neutron energy spectra, $\Phi(E)$, has to be unfolded from the pulse height spectra $dN/dH$ as a result of neutron-proton recoil in BC501A liquid scintillator.

The detector response $R(H, E)$, the differential pulse height spectrum $dN/dH$, and the neutron energy spectrum $\Phi(E)$, are related through the Fredholm integral equation of the first type as below [7] such that;

$$\frac{dN}{dH} = \int R(H,E)\Phi(E)dE \quad (1)$$

here, the radiation detector can directly give $dN/dH$, which is the result of folding the detector response functions and the energy distribution of the incident neutrons. When the spectrum is recorded by a multichannel analyzer, Eq. (1) takes the discrete form;

$$N_i = \sum_j R_{ij}\Phi_j \quad (2)$$

---

*Corresponding author (email: yonghao.t.chen@gmail.com)

where $N_i$ ($i=1, 2, …, n$) is the recorded counts in the $i$th channel, $\Phi_j$ ($j=1, 2, …, m$) is the radiation fluence in the $j$th energy interval, and $R_{ij}$ is the response matrix coupling the $i$th pulse height interval with the $j$th energy interval. Eq. (2) can be transformed into the matrix notation such that;

$$N = R\Phi \quad (3)$$

Where $N=(N_1, N_2, …, N_n)^T$, $\Phi=(\Phi_1, \Phi_2, …, \Phi_m)^T$, $R$ is the response matrix with size of $n \times m$.

Both Eq. (2) and (3) need to be inverted in order to obtain the neutron spectrum from the measured pulse height distribution, which can be seen as a mapping from the measured $n$-dimensional space of detector response to the $m$-dimensional space of neutron energy fluence. The task of determining the unknown $\Phi$ from the observable $N$ presents an ill-conditioned problem. To unfold the spectrum, several mathematical methods and computing algorithms have been used, such as least-squares [8], Monte Carlo Methods [9], genetic algorithm [10], and populated artificial neural networks (ANNs) in recently years [11-12]. These methods are able to unfold neutron energy spectra, however the accuracy and precision of the said methods are not as good as expected, particularly compared with the spectra provided by the International Organization for Standardization (ISO). Thus finding a robust method to unfold neutron spectra accurately is required.

The GRAVEL method is an iterative unfolding algorithm which was originally proposed by the Physikalisch Technische Bundesanstalt (PTB) for unfolding particle spectra from the measured pulse height distribution [13]. It has been successfully used for unfolding the spectra of γ-rays [14-15]. However, the application of GRAVEL method for unfolding the fast neutron spectra measured with liquid scintillation detector have not really been reported much more other than for PTB itself [16-17]. Herein the GRAVEL iterative algorithm is introduced in details, and a MatLab program based on GRAVEL method has been developed and successfully applied for unfolding the continuous neutron spectrum of $^{241}$Am-Be source and monoenergetic neutron spectrum of D-T source. The results exhibited good agreement with the ISO standard spectra.

This paper presents the overall procedure of measuring and unfolding the fast neutron energy spectra with BC501A liquid scintillation detector. First of all, the experiment setup for discriminating against γ-rays and measuring the recoil proton spectra of incident neutrons is described. Then the response matrix of the BC501A liquid scintillation detector used in our experiment is calculated by NRESP7 code. Finally, the unfolding results are presented.

## 2. Experiment

The experimental setup is shown in Fig. 1. A $\Phi 2''\times 2''$ cylindrical BC501A liquid scintillator coupled to a photomultiplier (PMT) with the tpye of 9807B of ET Enterprises with silicon oil was used to detect neutrons. In order to inhibit the scattering background of neutrons, the experiment was arranged in a spacious experimental hall (26.3 m in length, 11.4 m in width and 14 m in height), the neutron source ($^{241}$Am-Be source with the intensity of $2.5 \times 10^6$ n/s) and detector were put at the center of the hall. The detector was supported by a thin steel bracket and positioned perpendicularly to the ground, the distance from the scintillator front surface to the ground was 3.8 m. The neutron source was suspended 95 cm away from the scintillator front surface on its central axis. This type of arrangement could minimize the scattering neutron background.

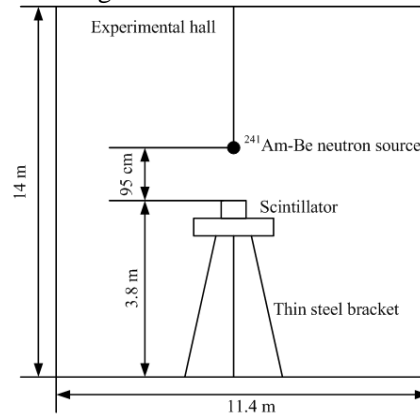
Figure 1 Experimental setup.

In order to evaluate the inhibition quality of scattering neutron background, this experiment setup was simulated by Monte Carlo N-Particle Transport Code (MCNP). It can be seen through the simulation result (Fig. 2) that this arrangement almost inhibited all the scattering neutrons, the scattering background only make a small contribution to the measured data thus the effect of background could be neglected.

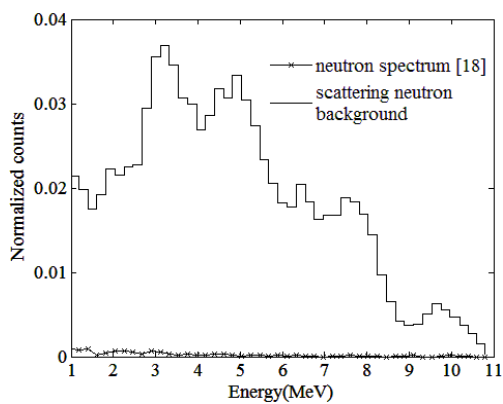

Figure 2 Simulation result of neutron scattering background for experiment setup with MCNP.

The block diagram in Fig. 3 shows the data acquisition system (DAQ) for measuring the recoil proton spectrum and discriminating against γ-rays with zero-crossing method.

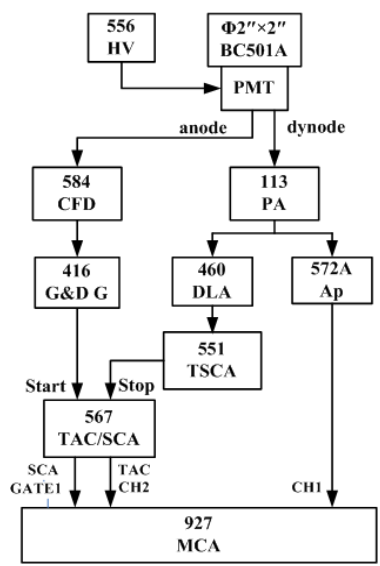

Figure 3 Block diagram of DAQ (all of the electronics are from ORTEC).

The DAQ system consisted of two branches: Ch1-for energy spectra measurement, and Ch2-for n-γ discrimination with zero-crossing method. The TAC output of 567 Time-to-Amplitude Converter/Single-Channel Analyzer (TAC/SCA) was the n-γ discrimination spectrum shown as Fig. 4. The neutron events in Fig. 4 were chosen by adjusting the SCA lower level of 567 TAC/SCA, namely tuned the SCA lower level to the valley channel between neutron and γ-ray peak (separating line in Fig. 4). The SCA output was sent to 927 as the gate signal for Ch1, then the recoil proton spectrum without γ-rays was obtained (Fig. 5). The measurement duration for $^{241}$Am-Be source was 60,399 s. The experiment details can be found in previous work [19].

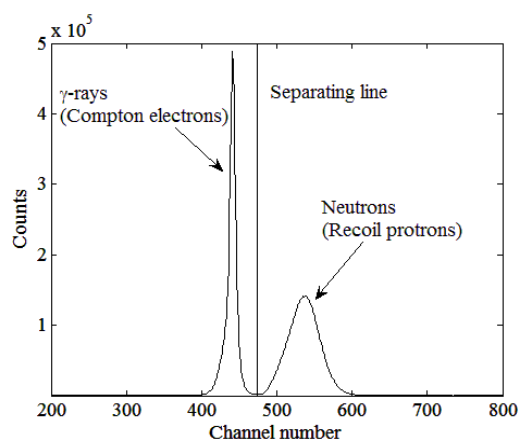

Figure 4 n- discrimination spectrum of $^{241}$Am-Be source.

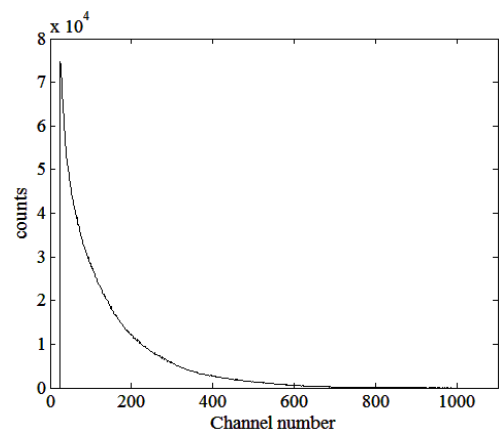

Figure 5 Recoil proton spectrum of $^{241}$Am-Be source.

The D-T source was measured with same DAQ system as shown in Fig. 3. However, the detector was placed in a hole on the shielding wall, and at the angle of 30 degree from the direction of deuterium ions beam with current of 70 μA and average energy of 135 keV. The distance from T target to detector was given as 13m and the duration 2,900 s.

## 3. Neutron spectra unfolding

### 3.1 Detector response matrix

As Eq. (1) exhibits, unfolding the measured pulse height spectra of neutrons requires a response matrix for the neutron energies in the energy range of interest. The response matrix can be determined experimentally or by calculations. In general, the response functions tend to be calculated by some specific program rather than measured experimentally because of various limited factors.

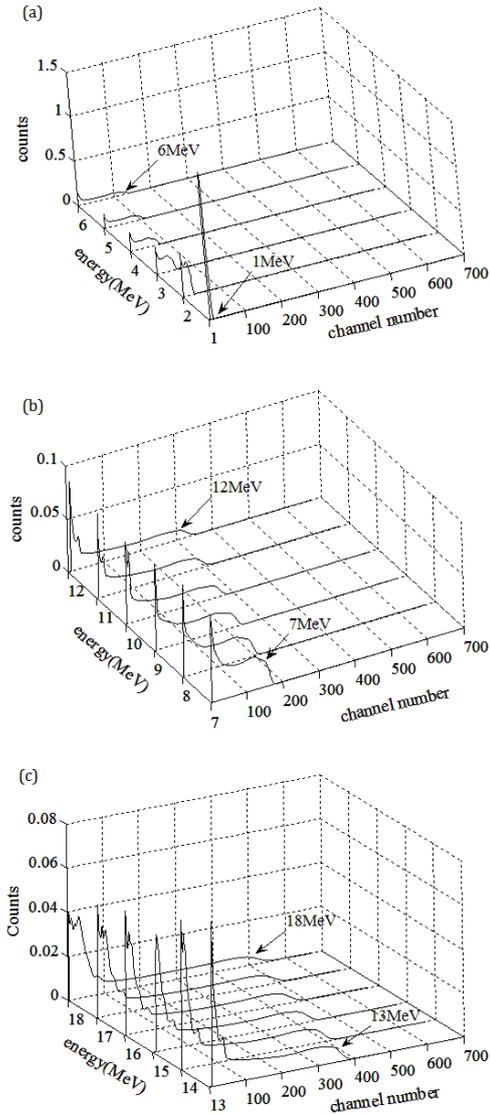

Figure 6 Response functions for $\Phi 2''\times 2''$ BC501A liquid scintillation detector from (a) 1MeV to 6MeV, (b) 7MeV to 12MeV, (c) 13MeV to 18MeV with 1MeV interval.

In this work, the response functions of $\Phi 2''\times 2''$ BC501A scintillation detector induced by monoenergetic neutrons were calculated by NRESP7 code [20]. The NRESP7 is a Monte Carlo code which calculates the detector response functions for NE213 scintillators (NE213 liquid scintillators are identical to BC501A scintillators in both design and chemical composition [21]) by incidence of fast monoenergetic neutrons in the energy range from 0.02 MeV to 20 MeV. An extensive set of cross section and differential cross section data has been included and various light output functions have been used in this code. It simulates all reactions of the neutrons inside or near the detector and calculates the light output induced by these reactions.

According to the neutron energy range of the $^{241}$Am-Be and D-T source, the response functions of BC501A scintillation detector in the range from 1MeV to 18MeV were calculated. There were 261 response functions. Fig. 6 shows response functions in the energy range from 1MeV to 18MeV with 1MeV interval.

### 3.2 GRAVEL unfolding method

GRAVEL is an iterative unfolding algorithm that uses a slight modification of the SAND-II algorithm. The iterative algorithm of GRAVEL is shown as below [13, 22];

$$\Phi_j^{K+1} = \Phi_j^K \exp\left(\frac{\sum_i W_{ij}^K \ln\left(\frac{N_i}{\sum_{j'} R_{ij}\Phi_{j'}^K}\right)}{\sum_i W_{ij}^K}\right) \quad (4)$$

where $N_i$ represents the measured counts in $i$th channel of pulse height spectrum. $R_{ij}$ is the response matrix coupling the $i$th pulse height interval with the $j$th energy interval, $W_{ij}$ is a weight factor defined as;

$$W_{ij}^K = \frac{R_{ij}\Phi_j^K}{\sum_{j'} R_{ij}\Phi_{j'}^K} \bullet \frac{N_i^2}{\sigma_i^2} \quad (5)$$

where $\sigma_i$ is the estimate of the measurement error, namely the square root of $N_i$. $i=1, \ldots, n$,

$j=1, \ldots, m$, with $n \sim m$.

The value of $\Phi_j^K$ is the neutron fluence in the $j$th energy interval after $K$th iteration. A first input spectrum, $\boldsymbol{\Phi}^0$, is needed when the iteration is started. A constant spectrum was used to initiate the iteration in this work [13].

GRAVEL is an iterative algorithm, so the question of stopping the iteration is a key point. The $\chi^2$ per degree of freedom, that is $\chi^2/n$, is used as the criterion for stopping the iteration.

The value of $\chi^2$ is defined as;

$$\chi^2 = \sum_i \frac{\left(\sum_j R_{ij}\Phi_j - N_i\right)^2}{\sigma_i^2} \qquad (6)$$

so the $\chi^2/n$ per degree of freedom, $\chi^2/n$, is given by;

$$\chi^2/n = \frac{1}{n}\sum_i \frac{\left(\sum_j R_{ij}\Phi_j - N_i\right)^2}{\sigma_i^2} \qquad (7)$$

It should be noted that every solution with $\chi^2/n \approx 1$ should be considered as consistent [13].

### 3.3 Results and Discussion

A MatLab unfolding iteration program was developed to carry out the GRAVEL algorithm and calculate the $\chi^2/n$ for each iteration based on the above principle.

The convergent curve of unfolding procedure for $^{241}$Am-Be neutrons is shown in Fig. 7. It can be seen that the final $\chi^2/n$ is almost to 1, which means the iteration is convergent.

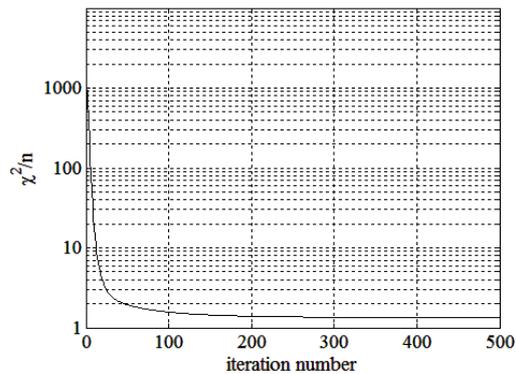

Figure 7 Convergent curve of unfolding procedure for $^{241}$Am-Be neutron source.

The $^{241}$Am-Be neutron spectra unfolded by GRAVEL method from its recoil proton spectrum (Fig. 5), unfolded by ANNs [23], taken from ISO [18] are shown in Fig. 8. The peaks of $^{241}$Am-Be neutron spectrum unfolded by GRAVEL are found approximately at 3.0, 5.0, 7.7, 10.0 MeV, which are in agreement with the reference that peaks of $^{241}$Am-Be neutron energy spectrum are observed near 3.1, 4.8, 7.7, 9.8 MeV [24].

It can also be seen from Fig. 8 that the GRAVEL unfolded result is in good agreement with the ISO [18] spectrum both in terms of the overall distribution and the position of peaks. The GRAVEL unfolded result is also more consistent than the results unfolded by Sharghi Ido [23] with the ANNs method. Our result appears to be more accurate comparatively.

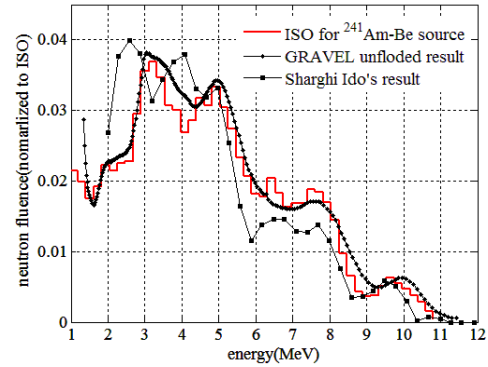

Figure 8 Comparison of the $^{241}$Am-Be neutron spectrum unfolded by GRAVEL method with the ISO reference and the Sharhi Ido's result unfolded by ANN.

The D-T neutrons were measured and unfolded under the same procedure. The final unfolded result is shown in Fig. 9.

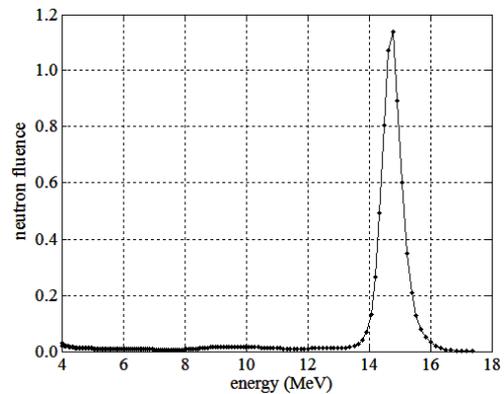

Figure 9 Neutron spectrum of D-T source unfolded by GRAVEL method.

The monoenergetic D-T neutrons were measured at the angle of 30 degree from the direction of deuterium ions beam (average energy is 135 keV) to detector, according to the data manual of accelerator neutron source, the energy of D-T neutrons in this direction is around 14.8MeV. The peak value in Fig.9 is about 14.78MeV, which is in good agreement with the data in the manual.

## 4. Conclusion

Herein, the overall procedure of unfolding the neutron energy spectra with a BC501A liquid scintillation detector was reported. Firstly, the experimental arrangement and electronic circuit for performing the n-γ discrimination and measuring the recoil proton spectra was introduced, and the experiment result of recoil proton spectrum of $^{241}$Am-Be neutron source was obtained. Then, the response matrix which was composed of 261 response functions for BC501A liquid scintillation detector was generated by Monte Carlo code NRESP7. Finally, combining the measured recoil proton method and simulated response matrix, the unfolding of the neutron spectra was carried out with the GRAVEL method.

A MatLab program based on the GRAVEL iterative algorithm has been developed and successfully applied for unfolding the fast neutron spectra. The continuous neutron spectrum of $^{241}$Am-Be source and monoenergetic neutron spectrum of D-T source were successfully unfolded by this MatLab program. The unfolded $^{241}$Am-Be neutron spectrum was in good agreement with the reference spectrum provided by ISO [18]. The GRAVEL unfolded spectrum of $^{241}$Am-Besource was more accurate compared with the reference published in recent years which is unfolded by ANNs [24]. The unfolded D-T neutron spectrum was in good agreement with the data in manual.

The original GRAVEL program developed by PTB is written with Fortran code [13]. It is a comparative large program which may not be workable under certain constrains. However, our developed program is a short code (200 lines). The complexity of the unfolding program is dramatically reduced.

*The authors would like to acknowledge Dr. Xiao Jun for the useful discussions on the unfolding algorithm, thank Zhu Chuanxin for the help on the experiment setup, as well as maintenance staff of the neutron generator in Institute of Nuclear Physics and Chemistry for providing monoenergetic D-T neutrons.*

1 Flaska M, Pozzi S A. Pulse-shape discrimination for identification of neutron sources using the BC-501A liquid scintillator. Proc. 8th Joint Int. Topical Meeting Mathematics & Computation and Supercomputing in Nuclear Applications (M&C+ SNA 2007). 15-19

2 Flaska M, Pozzi S A. Optimization of an offline pulse-shape discrimination technique for the liquid scintillator BC-501A. ORNL/TM-2006/120. Oak Ridge National Laboratory, Oak Ridge, Tennessee, 2006

3 Mullens J A, Edwards J D, Pozzi S A. Analysis of pulse-height for nuclear material identification. Institute of Nuclear Materials Management 45th Annual Meeting. Orlando, Florida, 2004

4 Guerrero C, Cano-Ott D, Fernandez-Ordonez M, et al. Analysis of the BC501A neutron detector signals using the true pulse shape. Nucl. Instrum. Meth. A, 2008, 597: 212-218

5 Marrone S, Cano-Ott D, Colonna N, et al. Pulse shape analysis of liquid scintillators for neutron studies. Nucl. Instrum. Meth. A, 2002, 490: 299-307

6 Chen Y H, Chen X M, Zhang X D, et al. Study of n-discrimination in low energy range (above 40 keVee) by charge comparison method with a BC501A liquid scintillation detector. Chinese Physics C. (Accepted)

7 Knoll G F. Radiation Detection and Measurement, fourth edition. Hoboken, New Jersey: John Wiley & Sons, Inc., 2010

8 Koohi-Fayegh, Green S, Scott M C. A comparison of neutron spectrum unfolding codes with a miniature NE213 detector. Nucl. Instrum. Meth. A, 2001, 460: 391-440

9 Sanna R, Obrien K. Monte-Carlo unfolding of neutron spectra. Nucl. Instrum. Meth., 1971, 91: 573-576

10 Mukherjee B. BONDI-97: a novel neutron energy spectrum unfolding tool using a genetic algorithm. Nucl. Instr. Meth. A, 1999, 432: 305-312

11 Vega-Carrillo H R, Hernandez-Davila V M, Manzanares-Acuna E, et al. Neutron spectrometry using artificial neural networks. Nucl. Instr. Meth. A, 2006, 41: 425-431

12 Avdic S, Pozzi S A, Protopopescu V. Detector response unfolding using artificial neural networks. Nucl. Instr. Meth. A, 2006, 565: 742-752

13 Matzke M. Unfolding of pulse height spectra: The

HEPRO program system. Report PTB-N-19, Physikalisch-Technische-Bundesanstalt, Braunschweig, 1994

14 Reginatto M, Goldhagen P, Neumann S. Spectrum unfolding, sensitivity analysis and propagation of uncertainties with the maximum entropy deconvolution code MAXED. Nucl. Instr. Meth. A, 2002, 476: 242-246

15 Kudo K, Takeda N, Koshikawa S, et al. Photon spectrometry in thermal neutron standard field. Nucl. Instr. Meth. A, 2002, 476: 213-217

16 Gressier V, Lacoste V, Lebreton L, et al. Characterisation of the IRSN CANEL/T400 facility producing realistic neutron fields for calibration and test purposes. Radiation protection dosimetry, 2004, 110(1-4): 523-527

17 Klein H, Neumann S. Neutron and photon spectrometry with liquid scintillation detectors in mixed field. Nucl. Instr. Meth. A, 2002, 476: 132-142

18 International Organization for Standardization. Reference neutron radiations-Part I: Characteristics and methods of production. ISO 8529-1, 2000

19 Chen Y H, Lei J R, Zhang X D, et al. Study of n-γ discrimination for 0.4-1MeV neutrons using the zero-crossing method with a BC501A liquid scintillation detector. Chinese Physics C, 2013, 37(4): 046202

20 Dietze G, Klein H. NRESP4 and NEFF4-Monte Carlo codes for the calculation of neutron response functions and detection efficiencies for NE213 scintillation detectors. PTB-ND-22, Braunschweig, 1982

21 Klein H. Neutron spectrometry in mixed field: NE213/BC501A liquid scintillation spectrometers. Radiat. Prot. Dosim., 2003, 107: 95-109

22 Matzke M. Unfolding of particle spectra. Proc. SPIE. 1997, 2867(59): 598-607

23 Sharghi Ido A, Bonyadi M R, Etaati G R, et al. Unfolding the neutron spectrum of a NE213 scintillator using artificial neural networks. App. Radia. and Isotop., 2009, 67: 1912-1918

24 Marsh J W, Thomas D J, Burke M. High resolution measurements of neutron energy spectra from Am-Be and Am-B sources. Nucl. Instrum. Meth. A, 1995, 366: 340-348